\input ./style/arxiv-general.cfg
\documentclass[aoas,MSNbibl,nameyear,seceqn,dvips]{arximspdf}
\makeatletter
   \@ifpackageloaded{graphicx}{}{\usepackage{graphicx}}
\makeatother
\usepackage{dcolumn}

\doi{10.1214/15-AOAS866}% Updated by VTEXPTS2LaTeX.exe, 07.10.2015 14:40
\volume{9}
\issue{4}
\pubyear{2015}
\firstpage{1709}
\lastpage{1725}
\docsubty{FLA}

\makeatletter
\newcolumntype{d}[1]{D{.}{.}{#1}}
\newcommand{\rrVert}{\Vert}
\newcommand{\llVert}{\Vert}
\newcommand{\argmin}{\mathop{\arg\min}\limits}
\def\mR{\mathbb{R}}
\makeatother

\begin{document}
\begin{frontmatter}

%\dochead{}
\title{Customized training with an application to mass spectrometric
imaging of cancer tissue}
\runtitle{Customized training}

\begin{aug}
% Corresponding author: Scott Powers - sspowers@stanford.edu% Updated by VTEXPTS2LaTeX.exe, 07.10.2015 15:47
%by VTEXPTS2LaTeX.exe, 07.10.2015 14:40
\author[A]{\fnms{Scott}~\snm{Powers}\corref{}\thanksref{T1}\ead[label=e1]{sspowers@stanford.edu}},
\author[A]{\fnms{Trevor}~\snm{Hastie}\thanksref{T2}\ead[label=e2]{hastie@stanford.edu}}
\and
\author[A]{\fnms{Robert}~\snm{Tibshirani}\thanksref{T3}\ead[label=e3]{tibs@stanford.edu}\ead[label=u1,url]{http://www.foo.com}}
\runauthor{S. Powers, T. Hastie and R. Tibshirani}
\affiliation{Stanford University}
%\dedicated{}
\address[A]{Department of Statistics\\
Stanford University\\
390 Serra Mall\\
Stanford, California 94305-4065\\
USA\\
\printead{e1}\\
\phantom{E-mail: }\printead*{e2}\\
\phantom{E-mail: }\printead*{e3}}
%\address[]{\\\printead{}}
\end{aug}
\thankstext{T1}{Supported by NSF Graduate Research Fellowship.}
\thankstext{T2}{Supported in part by NSF Grant DMS-14-07548 and NIH
Grant RO1-EB001988-15.}
\thankstext{T3}{Supported in part by NSF Grant DMS-99-71405 and NIH
Contract N01-HV-28183.}

% HISTORY:
%
\received{\smonth{4} \syear{2015}}% Updated by VTEXPTS2LaTeX.exe,
%07.10.2015 14:40
%
\revised{\smonth{7} \syear{2015}}% Updated by VTEXPTS2LaTeX.exe,
%07.10.2015 14:40

% ABSTRACT
%
\begin{abstract}
We introduce a simple, interpretable strategy for making predictions on test
data when the features of the test data are available at the time of model
fitting.
Our proposal---{\em customized training}---clusters the data to find training
points close to each test point and then fits an $\ell_1$-regularized model
(lasso) separately in each training cluster. This approach combines the local
adaptivity of $k$-nearest neighbors with the interpretability of the lasso.
Although we use the lasso for the model fitting, any supervised
learning method
can be applied to the customized training sets.
We apply the method to a mass-spectrometric imaging data set from an
ongoing collaboration in gastric
cancer detection which demonstrates the power and interpretability of the
technique. Our idea is simple but potentially useful in situations
where the
data have some underlying structure.
\end{abstract}

% KEYWORDS
% Pirmas kwd is didziosios raides
%
\begin{keyword}
\kwd{Transductive learning}
\kwd{local regression}
\kwd{classification}
\kwd{clustering}
\end{keyword}
\end{frontmatter}

\setcounter{footnote}{3}

%s1 #&#
\section{Introduction}

Recent advances in the field of personalized medicine have demonstrated the
potential for improved patient outcomes through tailoring medical
treatment to
the characteristics of the patient [\citet{HamburgCollins10}]. While these
characteristics most often come from genetic data, there exist other molecular
data on which to distinguish patients. In this paper we propose {\em customized
training}, a very general, simple and interpretable technique for local
regression and classification on large amounts of data in high
dimension. The
method can be applied to any supervised learning or transductive
learning task,
and it demonstrates value in applications to real-life data sets.

\begin{figure}

\includegraphics{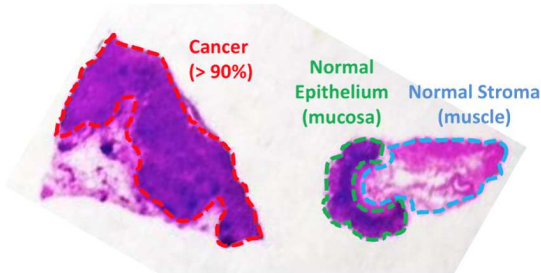}

\caption{Histopathological assessment of a banked tissue example. This
hematoxylin and eosin stain has been hand-labeled by a pathologist, marking
three regions: gastric adenocarcinoma (cancer), epithelium (normal) and stroma
(normal).}
\label{fig-histopathology}
\end{figure}

This paper is motivated by a newly proposed medical technique for inspecting
the edge of surgically resected tissue for the presence of gastric cancer
[\citet{Eberlin.etal14}]. Gastric is the second most lethal form of
cancer, behind
lung cancer [\citet{WHO13}], and the state-of-the-art treatment for
gastric cancer
is surgery to remove the malignant tissue. With this surgical procedure,
removal of {\it all} diseased tissue is critical to the prognosis for
the patient post-surgery. The new medical technique uses mass spectrometric
imaging, rather than visual inspection by a pathologist, to more
quickly and
more accurately evaluate the surgical margin of the tissue for the
presence of
cancerous cells. This new technique replaces the procedure wherein the tissue
samples are frozen until the pathologist is available to manually label the
tissue as cancer or normal (see Figure~\ref{fig-histopathology}).

%
%f1 #&#

\begin{figure}[t]

\includegraphics{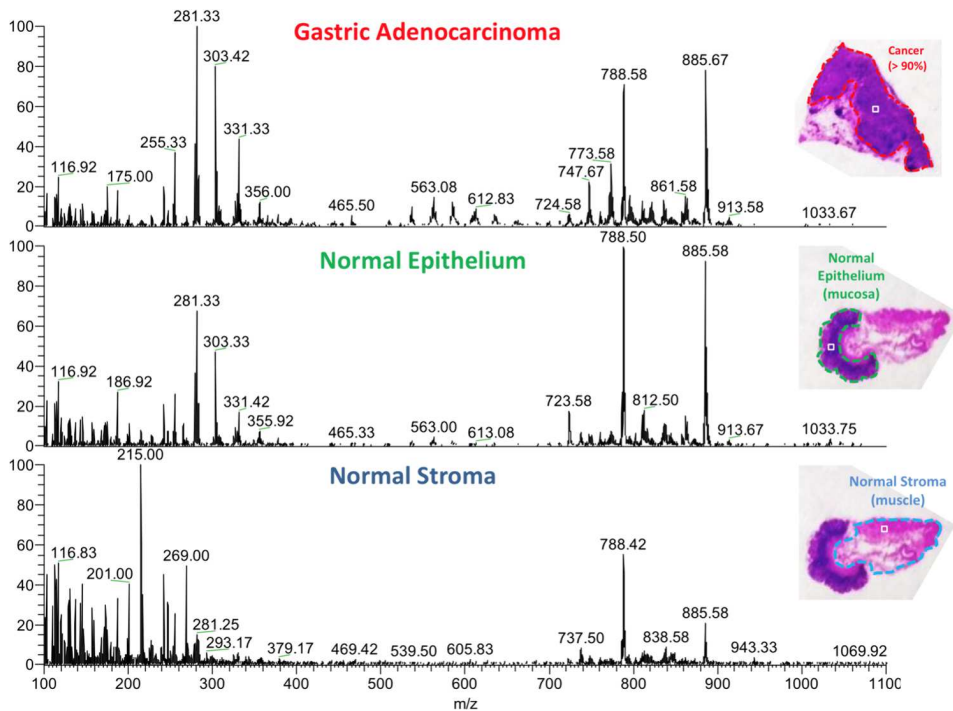}

\caption{DESI mass spectra for one pixel taken from each region in the banked
tissue example. The result of DESI mass spectrometric imaging is a 2D
ion image
with hundreds of pixels. Each pixel has an ion intensity measurement at
each of
thousands of mass-to-charge values, producing a mass spectrum. The three
mass spectra in the image correspond to one pixel each. The objective
is to
classify a pixel as cancer or normal on the basis of its mass spectrum.}
\label{fig-imaging}
\end{figure}

The data are images of surgical tissue from a desorption electrospray
ionization (DESI) mass spectrometer, which
records the abundance of ions at 13,320 mass-to-charge values at each of
hundreds of pixels. Hence, each data observation is a mass spectrum for a
pixel, as illustrated in Figure~\ref{fig-imaging}.

The 13,320 ion intensities from the mass spectrum for each pixel were averaged
across bins of
six\footnote{The third author's collaborators decided that six
was the appropriate bin size to reflect uncertainty in alignment due to
registration error.}
to yield 2220 features. Each pixel has been labeled by a
pathologist (after 2 weeks of sample testing) as epithelial, stromal or cancer,
the first two being normal tissue. Each of 20 patients contributed up
to three
samples, from some or all of the three classes. The training set
comprises 28
images from 14 patients, yielding 12,480 pixels, and the test set has
12 images
from 6 different patients, for a total of 5696 pixels.

%
%f2 #&#

In \citet{Eberlin.etal14} the authors use the lasso ($\ell_1$-regularized
multinomial regression) to model the probability that a pixel belongs
to each
of the three classes on the basis of the ion intensity in each bin of six
mass-to-charge values. In that study, the lasso performed favorably in
comparison with support vector machines and principal component
regression. For
a detailed description of the lasso, see Section~\ref{sub-lasso}. For the
purposes of the present paper, we collapse epithelial and stromal into one
class, ``Normal,'' and we adopt a loss function that assigns twice the penalty
to misclassifying a cancer cell as normal (false negative), relative to
misclassifying a normal cell as cancer (false positive). This loss function
reflects that missing a cancer cell is more harmful than making an
error in the
opposite direction. We collapse the two types of normal cells into one class
because our collaborators are interested in identifying only the cancer cells
for surgical resection. We find that treating epithelial and stromal as
separate classes does not meaningfully change our results.

The lasso classifier fit to the data from the 12,480 pixels
in the training set (with the regularization parameter $\lambda$
selected via
cross-validation; see Section~\ref{sub-cv}) achieves a
misclassification rate
of 2.97\% when used to predict the cancer/normal label of the 5696
pixels in
the test set. Among cancer pixels the test error rate is 0.79\%, and among
normal pixels the test error rate is 4.16\%. These results represent a
significant improvement over the subjective classifications made by
pathologists, which can be unreliable in up to 30\% of patients
[\citet{Eberlin.etal14}], but the present paper seeks to improve these results
further. By using customized training sets, our method fits a separate
classifier for each patient, creating a locally linear but globally nonlinear
decision boundary. This rich classifier leads to more accurate
classifications by using training data most relevant to each patient when
modeling his or her outcome probabilities.

%s1.1 #&#
\subsection{Transductive learning}

Customized training is best suited for the category of problems known in
machine learning literature as transductive learning, in contrast with
supervised learning or semi-supervised learning. In all of these problems,
both the dependent and the independent variables are observed in the training
data set (we say that the training set is ``labeled'') and the
objective is to
predict the dependent variable in a test data set.
The distinction between the three types of problems is
as follows: in supervised learning, the learner does not have access to the
independent variables in the test set at the time of model fitting,
whereas in
transductive learning the learner does have access to these data at model
fitting. Semi-supervised learning is similar in that the learner has
access to
unlabeled data in addition to the training set, but these additional
data do not belong to the test set on which the learner makes predictions.
Customized training leverages information in the test data by choosing
the most
relevant training data on which to build a model to make better predictions.
We have found no review of transductive learning techniques, but for a review
of techniques for the related semi-supervised problem, see \citet{Zhu07}.

In Section~\ref{sec-methods} we introduce customized training and
discuss related methods.
Section~\ref{sec-simulation} investigates the performance of customized
training and competing methods in a simulation study. Results on the motivating
gastric cancer data set are presented, with their interpretation, in
Section~\ref{sec-results}. We apply our method and others to a battery
of real data sets
from the UCI Machine Learning Repository in Section~\ref{sec-battery}. The
manuscript concludes with a discussion in Section~\ref{sec-discussion}.

%s2 #&#
\section{Customized training}
\label{sec-methods}

First we introduce some notation. The data we are given are
$X_{\mathrm{train}}$, $Y_{\mathrm{train}}$ and
$X_{\mathrm{test}}$. $X_{\mathrm{train}}$ is an
$n \times p$ matrix of predictor variables, and $Y_{\mathrm{train}}$
is an $n$-vector of response variables corresponding to the $n$ observations
represented by the rows of $X_{\mathrm{train}}$. These response
variables may be qualitative or quantitative.
$X_{\mathrm{test}}$ is an $m \times p$ matrix of the same $p$
predictor variables measured on $m$ test observations. The goal is to predict
the unobserved random $m$-vector $Y_{\mathrm{test}}$ of responses
corresponding to the observations in $X_{\mathrm{test}}$.

Let $\hat f_\Lambda(\cdot)$ denote the prediction made by some learning
algorithm, as a function of $X_{\mathrm{train}}$,
$Y_{\mathrm{train}}$, $X_{\mathrm{test}}$ and an ordered
set $\Lambda$ of tuning parameters. So
$\hat f_\Lambda(X_{\mathrm{train}}, Y_{\mathrm{train}},
X_{\mathrm{test}})$ is an $m$-vector. For qualitative
$Y_{\mathrm{train}}$, $\hat f_\Lambda$ is a classifier,\vspace*{1pt} while for
quantitative $Y_{\mathrm{train}}$, $\hat f_\Lambda$ fits a\vspace*{2pt}
regression. We evaluate the performance of $\hat f_\Lambda$ with
$L(\hat f_\Lambda(X_{\mathrm{train}}, Y_{\mathrm{train}},
X_{\mathrm{test}}), Y_{\mathrm{test}})$, where the loss
function $L$ is often taken to be, for example, the number of
misclassifications for a qualitative response, or squared error for a
quantitative response.

The customized training method partitions the test set into $G$ subsets and
fits a separate model $\hat f_\Lambda$ to make predictions for each
subset. In
particular, each subset of the test set uses only its own, ``customized''
subset of the training set to fit $\hat f_\Lambda$. Identifying subsets
of the
training data in this way leads to a model
that is locally linear but rich globally. Next, we propose two
methods for partitioning the test set and specifying the customized training
subsets.

%s2.1 #&#
\subsection{Clustering}
\label{sub-cluster}

Often test data have an inherent grouping structure, obviating the need to
identify clusters in the data using unsupervised learning techniques. Avoiding
clustering is
especially advantageous on large data sets for which it would be very expensive
computationally to cluster the data. For example, in the motivating application
for the present manuscript, test data are grouped by patient, so we avoid
clustering the 5696 test observations in 2220 dimensions by using patient
identity as the cluster membership for each test point.

Given the $G$ ``clusters'' identified by the grouping inherent to the test
data, we identify the customized training set for each test cluster as follows:
first, for each observation in the cluster, find the $R$ nearest
neighbors in
the training set to that observation, thus defining many
cardinality-$R$ sets
of training observations, one for each test point in the cluster.
Second, take
the union of these sets as the customized training set for the cluster.
So the
customized training set is the set of all training points that are one
of the
$R$ nearest neighbors of any test point in the cluster. $R$ is a tuning
parameter that could in principle be chosen by cross-validation, but
we have
found that $R = 10$ works well in practice and that results are not
particularly sensitive to this choice.

%
%f3 #&#
\begin{figure}
\includegraphics{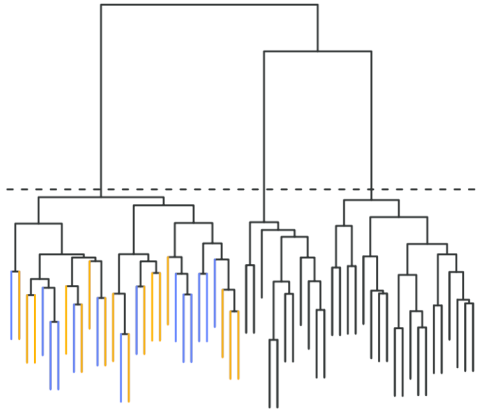}

\caption{A dendrogram depicting joint clustering of training and test
data, which is the method proposed for partitioning the test data and
identifying customized training sets when the test data have no inherent
grouping. Here the dendrogram is cut at a height to yield $G = 3$
clusters. Within the left cluster, the training data (blue leaves) are
used to fit the model and make predictions for the test data (orange leaves).}\vspace*{-3pt}
\label{fig-illustration-a}\label{fig3}
\end{figure}

When the test data show no inherent grouping, customized
training works by jointly clustering the training and test observations
according to their predictor variables. Any clustering method can be
used; here
we apply hierarchical clustering with complete linkage to the data
$(X^T_{\mathrm{train}}, X^T_{\mathrm{test}})^T$.
Then we
cut the dendrogram at some height $d_G$, producing $G$ clusters, as illustrated by Figure \ref{fig3}. In
each cluster we train a classifier on the training observations within that cluster.
This model is then used to make predictions for the test observations within
the cluster. In this case, $G$ is a tuning parameter to be chosen by
cross-validation (see Section~\ref{sub-cv}).

%s2.2 #&#
\subsection{Classification and regression}
\label{sub-lasso}

The key idea behind our method is the selection of a customized
training set
for each group in the test set. Once these individualized training sets are
identified, any supervised classification (or regression, in the case of
quantitative outcomes) technique can be used to fit $\hat f_\Lambda$
and make
predictions for the test
set. We suggest using $\ell_1$-regularized generalized linear models
because of
their interpretability. Customized training complicates the model by expanding
it into a compilation of $G$ linear models instead of just one. But using
$\ell_1$ regularization to produce sparse linear models conserves
interpretability. For an $n\times p$ predictor matrix $X$ and corresponding
response vector $y$, an $\ell_1$-regularized generalized linear model solves
the optimization problem
%
%
%e2.1 #&#
\begin{equation}
\label{eqn-glmnet} \min_{\beta_0, \beta\in\mR^p} - \frac{1}n\sum\ell(
\beta_0, \beta| x_i, y_i) + \lambda\llVert
\beta\rrVert _1,
\end{equation}
where $\ell(\cdot)$ here is the log-likelihood function and depends on the
assumed distribution of the response. For example, for linear
regression (which
we use for quantitative response variables),
\[
y_i|x_i,\beta_0,\beta\sim\operatorname{Normal}
\bigl(\beta_0 + \beta^Tx_i,
\sigma^2\bigr),
\]
while for logistic regression (which we use for binary response variabes),
\[
y_i|x_i,\beta_0,\beta\sim\operatorname{Binomial}
\biggl(1, \frac{e^{\beta_0 + \beta^T x_i}}{1 + e^{\beta_0 + \beta^T x_i}}\biggr).
\]
For multiclass qualitative response variables we use the multinomial
distribution in the same framework. The estimated regression
coefficient vector
$\hat\beta$ that solves the optimization problem (\ref{eqn-glmnet}) can be
interpreted as the contribution of each predictor to the distribution
of the
response, so by penalizing $\|\beta\|_1$ in (\ref{eqn-glmnet}), we encourage
solutions for which many entries in $\hat\beta$ are zero, thus simplifying
interpretation [\citet{Tibshirani96}].

Regardless of the $\hat f_\Lambda$ chosen, for $g = 1, \ldots, G$, let $n_k$
denote the number of observations in the customized training set for the
$k$th test cluster, and let\vspace*{1pt}
$X^k_{\mathrm{train}}$ denote the $n_k \times p$ submatrix of
$X_{\mathrm{train}}$ corresponding to these observations,
with $Y^k_{\mathrm{train}}$ denoting the corresponding responses.
Similarly, let $m_k$ denote the number of test observations in the
$k$th cluster, and let $X^k_{\mathrm{test}}$
denote the $m_k \times p$ submatrix of $X_{\mathrm{test}}$
corresponding to these training observations, with
$Y^k_{\mathrm{test}}$ denoting the corresponding responses.
Once we
have a partition of the test set into $G$ subsets (some of which may
contain no
test observations), with tuning parameter $\Lambda$ our prediction for
$Y^k_{\mathrm{test}}$ is
%
%
%e2.2 #&#
\begin{equation}
\label{eq-prediction} \hat Y^k_{\mathrm{test}} = \hat f_{\Lambda}
\bigl(X^k_{\mathrm{train}}, Y^k_{\mathrm{train}},
X^k_{\mathrm{test}}\bigr).
\end{equation}

Note that if joint clustering is used to partition the test data, the
customized training set for the $k$th test cluster may be
empty, in which case
$\hat f_{\Lambda}(X^k_{\mathrm{train}},
Y^k_{\mathrm{train}}, X^k_{\mathrm{test}})$ is undefined.
The problem is not frequent, but we offer in Section~\ref
{sub-rejections} one
way (of several) to handle it. Once we
have predictions for each subset, they are combined into the $m$-vector
$\mathrm{CT}_{G, \Lambda} (X_{\mathrm{train}},
Y_{\mathrm{train}}, X_{\mathrm{test}} )$,
which we
take as our prediction for~$Y_{\mathrm{test}}$.

%s2.3 #&#
\subsection{Cross-validation}
\label{sub-cv}

Because customized training reduces the training set for each test observation,
if the classification and regression models from Section~\ref{sub-lasso}
were not regularized, they would run the risk of overfitting the data. The
regularization parameter $\lambda$ in (\ref{eqn-glmnet}) must be large enough
to prevent overfitting but not so large as to overly bias the model
fit. This
choice is known as the bias-variance trade-off in statistical learning
literature [\citet{Hastie.etal09}].

The number of clusters $G$ is also a tuning parameter that controls the
flexibility of the model. Increasing $G$ reduces the bias of the model
fit, while decreasing $G$ reduces the variance of the model fit. To determine
the optimal values of $G$ and~$\Lambda$, we use standard
cross-validation to
strike a balance between bias and variance. Because transductive
methods have
access to test features at training time, we explain carefully in this section
what we mean by standard cross-validation.

The training data are randomly partitioned into $J$ approximately equal-sized
folds (typically $J = 10$). For
$j = 1, \ldots, J$, $X^{(j)}_{\mathrm{train}}$ denotes the\vspace*{2pt}
submatrix of
$X_{\mathrm{train}}$ corresponding to the data in the $j$th fold,
and $X^{(-j)}_{\mathrm{train}}$ denotes the submatrix of data
{\em not} in the $j$th fold. Similarly, $Y^{(j)}_{\mathrm{train}}$
denotes the responses corresponding to the data in the $j$th fold, and
$Y^{(-j)}_{\mathrm{train}}$ denotes responses {\em not} in the
$j$th fold.

We consider $\mathcal G$ and $A$ as the sets of possible values for $G$ and
$\Lambda$, respectively. In practice, we use $\mathcal G = \{1, 2, 3,
5, 10\}$.
We search over the grid $\mathcal G \times A$, and the CV-selected parameters
$G$ and $\Lambda$ are
\[
\bigl(G^*, \Lambda^*\bigr) = \argmin_{G \in\mathcal G, \Lambda\in A}\sum
_{j = 1}^J L \bigl(\mathrm{CT}_{G, \Lambda}
\bigl(X^{(-j)}_{\mathrm{train}}, Y^{(-j)}_{\mathrm{train}},
X^{(j)}_{\mathrm{train}} \bigr), Y^{(j)}_{\mathrm{train}} \bigr).
\]

In more detail, the $G$ clusters for
$\mathrm{CT}_{G, \Lambda}(X^{(-j)}_{\mathrm{train}},
Y^{(-j)}_{\mathrm{train}}, X^{(j)}_{\mathrm{train}})$ are
obtained as described in Section~\ref{sub-cluster}, and the loss for the
$j$th fold is given by
\begin{eqnarray*}
&& L \bigl(\mathrm{CT}_{G, \Lambda} \bigl(X^{(-j)}_{\mathrm{train}},
Y^{(-j)}_{\mathrm{train}}, X^{(j)}_{\mathrm{train}} \bigr),
Y^{(j)}_{\mathrm{train}} \bigr)
\\
&&\qquad =\sum_{k = 1}^GL \bigl(\hat
f_\Lambda \bigl(X^{(-j)^k}_{\mathrm{train}}, Y^{(-j)^k}_{\mathrm{train}},
X^{(j)^k}_{\mathrm{train}} \bigr), Y^{(j)^k}_{\mathrm{train}} \bigr).
\end{eqnarray*}

%s2.4 #&#
\subsection{Out-of-sample rejections}
\label{sub-rejections}

As noted in Section~\ref{sec-methods}, when joint clustering is used to
partition the test data and identify customized training sets,
predictions for
a particular test subset may be undefined because the corresponding customized
training subsets do not contain any observations. Using the convention of
\citet{BottouVapnik92}, we refer to this event as a {\em
rejection} (although it
might be more naturally deemed an {\em abstention}). The number of rejections,
then, is the number of test observations for which our procedure fails
to make
a prediction due to an insufficient number of observations in the customized
training set.

Typically, in the machine learning literature, a rejection occurs when a
classifier is not confident in a prediction, but that is not the case
here. For
customized training, a rejection occurs when there are no training observations
close to the observations in the test set. This latter problem has not often
been addressed in the literature [\citet{BottouVapnik92}]. Because the
test data
lie in a region of the feature space poorly represented in the training
data, a classifier might make a very confident, incorrect prediction.

We view the potential for rejections as a virtue of the method,
identifying situations in which it is best to make no prediction at all because
the test data are out-of-sample, a rare feature for machine learning
algorithms. In practice, we observe that rejections are rare;
Table~\ref{tab-rejections} gives a summary of all rejections in the
battery of
machine learning data sets from Section~\ref{sec-battery}.

If a prediction must be made, there are many ways to get around
rejections. We
propose simply cutting the dendrogram at a greater height $d^\prime>
d_G$ so
that the test cluster on which the rejections occurred is combined with another
test cluster until the joint customized training set is large enough to make
predictions. Specifically, we consider the smallest $d^\prime$ for
which the
predictions are defined. Note that we update the predictions only for the
test observations on which the method previously abstained.

%s2.5 #&#
\subsection{Related work}

Local learning in the transductive setting has been proposed before
[\citet{Zhou.etal04,WuSchoelkopf07}]. There are other related methods as well,
for example, transductive regression with elements of local learning
[\citet{CortesMohri07}] or local learning that could be adapted to the
transductive setting [\citet{Yu.etal09}]. The main contribution of this paper
relative to previous work is the simplicity and interpretability of customized
training. By combining only a few sparse models, customized training
leads to a
much more parsimonious model than other local learning algorithms, easily
explained and interpreted by subject-area scientists.

More recently, local
learning has come into use in the transductive setting in applications related
to personalized medicine. The most relevant example to this paper is evolving
connectionist systems [\citet{Ma12}], but again our proposal for customized
training leads to a more parsimonious and interpretable model. Personalized
medicine is an exciting area of potential application for customized training.

Several methods [\citet{GuHan13,LadickyTorr11,TorgoDaCosta03}]
similarly partition the feature space and fit separate classification or
regression models in each region. However, in addition to lacking the
interpretability of our method, these techniques apply only to the supervised
setting and do not leverage the additional information in the transductive
setting. Others have approached a similar problem
using mixture models [\citet{Fu.etal10,ShahbabaNeal09,Zhu.etal11}], but these
methods also come with a great computational burden, especially those
which use
Gibbs sampling to fit the model instead of an EM algorithm or variational
methods.

Variants of customized training could also be applied in the supervised and
semi-supervised setting. The method would be semi-supervised if instead
of test
data other unlabeled data were used for clustering and determining the
customized training set for each cluster. The classifier or regression obtained
could be used to make predictions for unseen test data by assigning
each test
point to a cluster and using the corresponding model. A supervised
version of
customized training would cluster only the training data and fit a
model for
each cluster using the training data in that cluster. Again,
predictions for
unseen test data could be made after assigning each test point to one
of these
clusters. This approach would be similar to \citet{JordanJacobs94}.

%s2.5.1 #&#
\subsubsection{Alternative methods}
\label{sub-alternative}

To compare customized training against the state of the art, we apply five
other machine learning methods to the data sets in Sections~\ref
{sec-simulation}, \ref{sec-results} and \ref{sec-battery}.
\begin{enumerate}[KSVM]
\item[ST] Standard training. This method uses the $\ell_1$-penalized regression
techniques outlined in Section~\ref{sub-lasso}, training one model on all
of the training set. The regularization parameter $\lambda$ is
chosen through cross-validation.
\item[SVM] Support vector machine. The cost-tuning parameter is chosen through
cross-validation.
\item[KSVM] $K$-means $+$ SVM. We cluster the training data into $K$
clusters via
the $K$-means algorithm and fit an SVM to each training cluster. Test data
are assigned to the nearest cluster centroid. This method is a simpler,
special case of the clustered SVMs proposed by \citet{GuHan13},
whose recommendation of $K = 8$ we use.
\item[RF] Random forests. At each split we consider $\sqrt{p}$ of the $p$
predictor variables (classification) or $p/3$ of the $p$ predictor
variables (regression).
\item[KNN] $k$-nearest neighbors. This simple technique for classification
and regression contrasts the performance of customized training with
another ``local'' method. The parameter $k$ is chosen via cross-validation.
\end{enumerate}

%s3 #&#
\section{Simulation study}
\label{sec-simulation}

We designed a simulation to demonstrate that customized training improves
substantially on standard training in situations where one would expect
it to
do so: when the data belong to several clusters, each with a different
relationship between features and responses. We consider real-valued responses
(a regression problem) for the sake of variety. We simulated $n$ training
observations and $m$ test observations in $p$ dimensions, each observation
belonging to one of $3$ classes. The frequency of the $3$ classes was
determined by a Dirichlet$(2, 2, 2)$ random variable. The centers
$c_1, c_2, c_3$ of the $3$ classes were generated as i.i.d. $p$-dimensional
normal random variables with covariance $\sigma_c^2I_p$.

Given the class membership $z_i \in\{1, 2, 3\}$ of the $i$th observation,
$x_i$ was generated from a normal distribution with mean $c_{z_i}$ and
covariance matrix $I_p$. The coefficient vector $\beta^k$ corresponding
to the
$k$th class had $p/10$ entries equal to one, with the rest being zero,
reflecting a sparse coefficient vector. The nonzero entries of $\beta
^k$ were
sampled uniformly at random, independently for each class $k$. Given
the class
membership $z_i$ and coefficient vector $x_i$ of the $i$th observation, the
response $y_i$ had a normal distribution with mean $(\beta^{z_i})^Tx_i$ and
standard deviation one.

%
%f4 #&#
\begin{figure}

\includegraphics{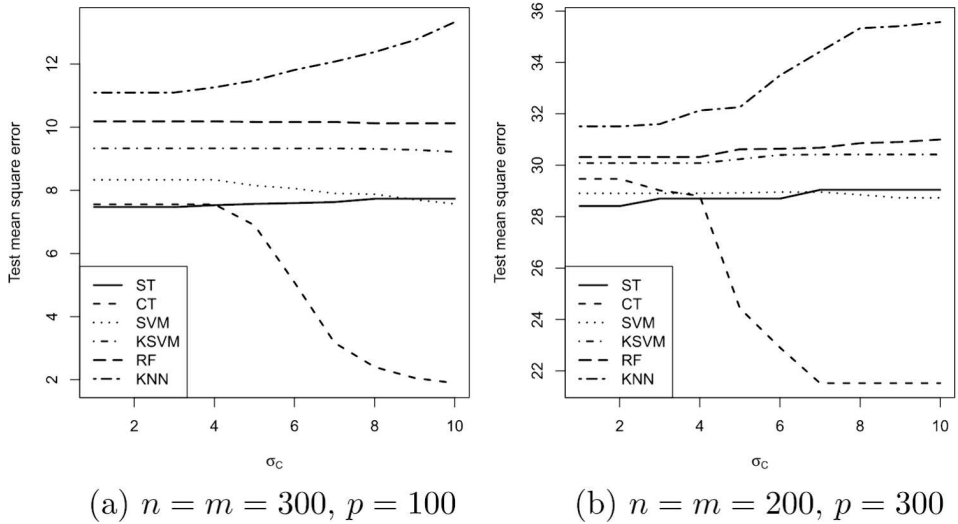}

% \caption{$n = m = 300$, $p = 100$}
% \caption{$n = m = 200$, $p = 300$}
\caption{Simulation results. In \textup{(a)}, the low-dimensional setting, as
$\sigma_C$ increases and the clusters separate, the test error for
customized training drops, while the
test error for other methods remains high. In \textup{(b)}, the test errors are
much larger overall, but the same pattern persists: customized training
leads to improved results as the clusters separate.}
\label{fig-simulation}
\end{figure}

We conduct two simulations, the first with $n = m = 300$, $p = 100$
(the low-dimensional setting), and the
second with $n = m = 200$, $p = 300$ (the high-dimensional setting).
In each case, we vary $\sigma_C$ from 0 to 10. Figure~\ref
{fig-simulation} shows the results. We observe that in both settings,
customized training leads to significant improvement in test mean
square error
as the clusters separate (increasing $\sigma_C$). In the high-dimensional
setting, the errors are expectedly much larger, but the same pattern is
evident. For KSVM in this simulation we fix $K = 3$, thus cheating and giving
the algorithm the number of clusters, whereas customized training
learns the
number of clusters from the data. For this reason, the performance of
KSVM does
not improve as the clusters separate. In fact, it is because none of
the other
methods make an attempt to identify the number of clusters that they do not
improve as the clusters separate.

%s4 #&#
\section{Results on gastric cancer data set}
\label{sec-results}

We applied customized training to the mass-spectrometric imaging data
set of
gastric cancer surgical resection margins with the goal of improving on the
results obtained by standard training.
As described in Section~\ref{sub-cluster}, we obtained a customized training
set for each test patient by finding the 10 nearest neighbors of each pixel
in that patient's images and using the union of these nearest-neighbor sets.
Table~\ref{tab-source} shows from which training patients the customized
training set came, for each test patient. The patient labels have been ordered
to make the structure in these data apparent: test patients 1--3 rely heavily
on training patients 1--7 for their customized training sets, while test
patients 4--6 rely heavily on training patients 9--14 for their customized
training sets.

%
%t1 #&#
\begin{table}%[h]
\tabcolsep=0pt
\caption{Source patients
used in customized training sets for six test patients.
Each column shows,
for the corresponding test patient, what percentage of observations in that
patient's customized training set came from each of the training patients.
Patient labels have been permuted to show the structure in the data: test
patients 1--3 get most of their training sets from patients 1--7,
while test patients 4--6 get most of their training sets from patients 9--14}\label{tab-source}
\begin{tabular*}{\tablewidth}{@{\extracolsep{\fill}}@{}lcd{2.1}d{2.1}d{2.1}d{2.1}d{2.1}d{2.1}d{2.1}d{2.1}d{2.1}d{2.1}d{2.1}d{2.1}d{2.1}d{2.1}@{}}
\hline
&& \multicolumn{14}{c@{}}{\textbf{Training  patient}}\\[-6pt]
&& \multicolumn{14}{c@{}}{\hrulefill}\\
&& \multicolumn{1}{c}{\textbf{1}} & \multicolumn{1}{c}{\textbf{2}} & \multicolumn{1}{c}{\textbf{3}}
& \multicolumn{1}{c}{\textbf{4}} & \multicolumn{1}{c}{\textbf{5}} & \multicolumn{1}{c}{\textbf{6}}
& \multicolumn{1}{c}{\textbf{7}} & \multicolumn{1}{c}{\textbf{8}} & \multicolumn{1}{c}{\textbf{9}}
& \multicolumn{1}{c}{\textbf{10}} & \multicolumn{1}{c}{\textbf{11}} & \multicolumn{1}{c}{\textbf{12}}
& \multicolumn{1}{c}{\textbf{13}} & \multicolumn{1}{c@{}}{\textbf{14}}\\
\hline
Test &1 & 41.7 & 39.5 & 22.8 & 4.4 & 20.8 & 4.2 & 3.5 & \multicolumn{1}{c}{--} & \multicolumn{1}{c}{--} & \multicolumn{1}{c}{--} & \multicolumn{1}{c}{--} & 0.3 & \multicolumn{1}{c}{--} & \multicolumn{1}{c@{}}{--} \\
patient &2 & 46.2 & 50.2 & 40.1 & 59.6 & 39.1 & 44.3 & 33.7 & \multicolumn{1}{c}{--} & \multicolumn{1}{c}{--} & 1.3 & \multicolumn{1}{c}{--} & 0.6 & \multicolumn{1}{c}{--} & 0.9 \\
& 3 & 12.0 & 9.5 & 36.9 & 34.5 & 28.4 & 50.8 & 61.1 & \multicolumn{1}{c}{--} & 38.1 & \multicolumn{1}{c}{--} & 6.6 & 32.8 & 3.6 & 8.2 \\
& 4 & \multicolumn{1}{c}{--} & 0.2 & \multicolumn{1}{c}{--} & 1.2 & \multicolumn{1}{c}{--} & \multicolumn{1}{c}{--} & 0.7 & 38.1 & 55.8 & 73.7 & 20.4 & 21.2 & 6.6 & 25.4 \\
&5 & \multicolumn{1}{c}{--} & \multicolumn{1}{c}{--} & \multicolumn{1}{c}{--} & 0.2 & \multicolumn{1}{c}{--} & 0.6 & 0.3 & 52.4 & 2.2 & 7.7 & 50.9 & 19.7 & 52.9 & 25.3 \\
&6 & \multicolumn{1}{c}{--} & 0.7 & 0.2 & 0.2 & 11.6 & \multicolumn{1}{c}{--} & 0.7 & 9.5 & 3.9 & 17.3 & 22.2 & 25.3 & 36.9 & 40.1 \\
\hline
\end{tabular*}
\end{table}

In this setting it is more harmful to misclassify cancer tissue as
normal than
it is to make the opposite error, so we chose to use a loss function that
penalizes a false negative (labeling a cancer pixel as normal) twice as much
as it does a false positive (labeling a normal pixel as cancer). We observe
that the results are not sensitive to the choice of the loss function
(in the
range of penalizing false negatives equally to five times as much as false
positives) in terms of comparing customized training against standard training.
We compare the results of customized training against the results of standard
training for $\ell_1$-regularized binomial regression---the method used by
\citet{Eberlin.etal14}---in Table~\ref{tab-results}.

%t2 #&#
\begin{table}[b]
\tabcolsep=0pt
\caption{Error rates for customized training and standard training on the
gastric cancer test data, split by patient and true label of the pixel (cancer
or normal), with the lower overall error rate for each patient in bold. Each
error rate is expressed by the percentage of pixels
misclassified. Customized training leads to slightly higher errors for patients
3 and 4 but much lower errors for all other
patients and roughly half the error rate overall}\label{tab-results}
\begin{tabular*}{\tablewidth}{@{\extracolsep{\fill}}@{}lcccccccc@{}}
\hline
\multicolumn{2}{l}{\textbf{Test patient}} & \multicolumn{1}{c}{\textbf{1}} & \textbf{2} & \textbf{3} & \textbf{4} & \textbf{5} & \textbf{6} & \textbf{All}\\
\hline
Standard& Cancer & \multicolumn{1}{c}{--} & 2.67 & 0.21 & \multicolumn{1}{c}{--} & \multicolumn{1}{c}{--} & 2.70 & 1.54 \\
training & Normal & 13.60 & 0.81 & 0.13 & 0.00 & 6.37 & 3.63 & 3.78 \\
(lasso) & Overall & 13.60 & 1.61 & \bf{0.18} & \bf{0.00} & 6.37 & 3.14 & 2.98
\\[3pt]
Customized& Cancer & \multicolumn{1}{c}{--} & 1.07 & 0.11 & \multicolumn{1}{c}{--} & \multicolumn{1}{c}{--} & 1.80 & 0.74 \\
training & Normal & \phantom{0}8.66 & 0.00 & 1.44 & 0.40 & 0.82 & 0.66 & 2.04 \\
(6-CT) & Overall & \phantom{0}$\mathbf{8.66}$ & \bf{0.46} & 0.71 & 0.40 & \bf{0.82} & \bf{1.26} & \bf{1.58} \\
\hline
\end{tabular*}
\end{table}

%
%t3 #&#
\begin{table}
\tabcolsep=0pt
\caption{Overall test error rates and run times for customized training
and the five other methods described in Section~\protect\ref{sub-alternative}}\label{tab-runtime}
\begin{tabular*}{\tablewidth}{@{\extracolsep{\fill}}@{}lcccccc@{}}
\hline
\textbf{Method} & \textbf{ST} & \textbf{CT} & \textbf{KSVM} & \textbf{KNN} & \textbf{RF} & \textbf{SVM}\\
\hline
Misclassification rate& 3.05\% & 1.58\% & 9.78\% & 9.18\% & 2.44\% & 2.07\%\\
Run time (minutes) & 2.1 & 2.4 & 6.0 & 7.6 & 21.9 & 197.8\\
\hline
\end{tabular*}
\end{table}

We observe that customized training leads to a considerable improvement in
results. For test patients 3 and 4, the test error is slightly higher for
customized training than for standard training, but for all other patients,
the test error for customized training is much lower. Overall, customized
training cuts the number of misclassifications in half from the results of
standard training. We focus on the comparison between customized and standard
training because they are the fastest methods to apply to this large
data set,
but, indeed, the other methods described in Section~\ref
{sub-alternative} are
also applicable. We report the overall test misclassification error and
the run
time for all methods in Table~\ref{tab-runtime}.

%s4.1 #&#
\subsection{Interpretation}

\begin{figure}[b]

\includegraphics{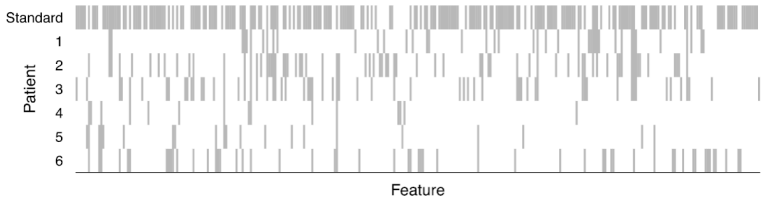}

\caption{Features selected by customized training for each patient
(variables not selected by any model are omitted from the $x$-axis). The
first row shows features selected via standard training. Visual inspection
suggests that patients 1, 2 and 3
have similar profiles of selected variables, whereas patients 4 and 5 have
selected-feature profiles that are more similar to each other than to other
patients. Using hierarchical clustering with Jaccard distance between the
sets of selected features to split the patients into two clusters, patients
1, 2 and 3 were in one cluster, with patients 4, 5 and 6 in the other.}
\label{fig-gastric}
\end{figure}

A key draw for customized training is that, although the decision
boundary is
more flexible than a linear one, interpretability of the fit is preserved
because of the sparsity of the model. In this example, there are 2220 features
in the data set, but the numbers of features selected for test patients 1
through 6 are, respectively, 42, 71, 62, 15, 21 and 54. Figure~\ref
{fig-gastric} shows which features are used in each patient's model, along
with the features used in the overall model with standard training.

%
%f5 #&#

We observe that some pairs of patients have more similar profiles of selected
features than other pairs of patients. For example, about 36\% of the features
selected for test patient 1 are also selected for test patient 2. And about
39\% of the features selected for test patient 3 are also selected for test
patient 2. This result is not surprising because test patients 1
through 3 take
much
of their customized training sets from the same training patients, as observed
above. Similarly, about 40\% of the features selected for patient 4 are also
selected for patient 6, and about 38\% of the features selected for
patient 5
are also selected for patient 6.

The third author's subject-area collaborators have suggested
that these data may actually suggest two subclasses of cancer; given that
customized training identifies two different groups of
models for predicting cancer presence, this subject-area knowledge
leads to a
potentially interesting interpretation of the results.

%s5 #&#
\section{Additional applications}
\label{sec-battery}

To investigate the value of customized training in practice, we applied
customized training (and the alternative methods from Section~\ref
{sub-alternative}) to a battery of classification data sets from the
UC Irvine Machine Learning Repository [\citet{UCI,Gil.etal12,Tsanas.etal13,Little.etal07,Mansouri.etal13,Kahraman.etal13}].
The data sets, listed in Table \ref{tab6}, were selected
not randomly but somewhat arbitrarily, covering a wide array of applications
and values of $n$ and~$p$, with a bias toward recent data sets. In Table \ref{tab4} we present
results on all 16 data sets to which the methods were applied, not just those
on which customized training performed well.

%
%t6 #&#
\begin{table}[b]
\tabcolsep=0pt
\caption{Data sets from UCI Machine Learning Repository [\citet{UCI}] used in Section~\protect\ref{sec-battery}}
\label{tab-abbreviations}\label{tab6}
\begin{tabular*}{\tablewidth}{@{\extracolsep{\fill}}@{}lccc@{}}
\hline
\textbf{Abbrv.} & \textbf{Data set name} & \textbf{Abbrv.} & \textbf{Data set name}\\
\hline
BS & Balance scale &
BCW & Breast cancer Wisconsin (diagnostic)\\
C & Chess (king-rook vs king-pawn) &
CMC & Contraceptive method choice\\
F & Fertility &
FOTP & First-order theorem proving\\
LSVT & LSVT voice rehabilitation &
M & Mushroom\\
ORHD & Optical recognition of handwritten digits &
P & Parkinsons\\
QSAR & QSAR biodegration &
S & Seeds\\
SPF & Steel plates faults &
TAE & Teaching assistant evaluation\\
UKM & User knowledge modeling &
V & Vowel\\
\hline
\end{tabular*}
\end{table}

%
%t4 #&#
\begin{table}[t]
\tabcolsep=0pt
\caption{Test error of customized training and the five other methods described
in Section~\protect\ref{sub-alternative} on 16 benchmark
data sets. The bold text indicates the best performance for each data set.
Customized training is competitive with the other methods and improves on
standard training more often than not}\label{tab-battery}\label{tab4}
\begin{tabular*}{\tablewidth}{@{\extracolsep{\fill}}@{}ld{4.0}d{3.0}ccd{2.0}ccccd{2.0}d{3.1}@{}}
\hline
\multicolumn{3}{c}{} & \multicolumn{1}{c}{\textbf{ST}} & \multicolumn{2}{c}{\textbf{CT}} & \multicolumn{1}{c}{\textbf{SVM}} & \multicolumn{1}{c}{\textbf{KSVM}} &
\multicolumn{1}{c}{\textbf{RF}} & \multicolumn{2}{c@{}}{\textbf{KNN}} \\[-6pt]
\multicolumn{3}{c}{} & \multicolumn{1}{c}{\hrulefill} & \multicolumn{2}{c}{\hrulefill} & \multicolumn{1}{c}{\hrulefill} & \multicolumn{1}{c}{\hrulefill} & \multicolumn{1}{c}{\hrulefill} & \multicolumn{2}{c@{}}{\hrulefill} \\
\textbf{Data} & \multicolumn{1}{c}{$\bolds{n}$} & \multicolumn{1}{c}{$\bolds{p}$} & \multicolumn{1}{c}{\textbf{Error}} & \multicolumn{1}{c}{\textbf{Error}}
& \multicolumn{1}{c}{$\bolds{G}$} & \multicolumn{1}{c}{\textbf{Error}} & \multicolumn{1}{c}{\textbf{Error}} & \multicolumn{1}{c}{\textbf{Error}} & \multicolumn{1}{c}{\textbf{Error}} &
\multicolumn{1}{c}{$\bolds{k}$} & \multicolumn{1}{c@{}}{\textbf{\%Imp\tabnoteref{TT1}}}\\
\hline
% latex table generated in R 3.1.3 by xtable 1.7-1 package
% Thu Apr 16 14:05:37 2015
BS & 313 & 4 &0.112 &0.099 & 3 & {\bf 0.086} &0.131 &0.131 &0.105 & 20 & 11.4  \\
BCW & 285 & 30 & {\bf 0.028} &0.035 & 2 &0.035 &0.038 & {\bf 0.028} &0.056 & 63 & -25  \\
C & 1598 & 38 &0.026 &0.021 & 10 &0.029 &0.046 & {\bf 0.006} &0.085 & 36 & 18.5  \\
CMC & 737 & 18 &0.485 & {\bf0.440} & 5 &0.479 &0.523 &0.472 &0.523 & 32 & 9.2  \\
F & 50 & 9 & {\bf0.160} & {\bf0.160} & 1 & {\bf0.160} & {\bf0.160} &0.180 &0.180 & 2 & \multicolumn{1}{c}{--} \\
FTP & 3059 & 51 &0.557 &0.530 & 5 &0.489 &0.444 & {\bf0.427} &0.508 & 47 & 4.7  \\
LSVT & 63 & 310 &0.126 &0.142 & 1 &0.111 &0.365 & {\bf 0.095} &0.222 & 15 & -12.5  \\
M & 4062 & 96 & {\bf 0.000} & {\bf 0.000} & 1 &0.001 &0.001 & {\bf 0.000} &0.001 & 15 & \multicolumn{1}{c}{--} \\
ORHD & 3823 & 62 &0.046 &0.043 & 2 &0.032 &0.049 & {\bf 0.027} &0.055 &
38 & 6.0  \\
P & 98 & 22 &0.268 &0.144 & 3 &0.154 &0.144 & {\bf 0.082} &0.123 & 5 &
46.1  \\
Q & 528 & 41 &0.176 & {\bf0.134} & 5 &0.146 &0.148 &0.140 &0.144 & 19
& 23.6  \\
S & 105 & 7 & {\bf 0.047} & {\bf 0.047} & 2 &0.066 &0.114 &0.104 &0.066
& 9 & \multicolumn{1}{c}{--} \\
SPF & 971 & 27 &0.321 &0.278 & 5 &0.273 &0.281 & {\bf0.246} &0.357 &
57 & 13.3  \\
TAE & 76 & 53 &0.720 & {\bf0.470} & 10 &0.653 &0.613 &0.506 &0.493 & 1
& 34.6  \\
UKM & 258 & 5 &0.041 & {\bf 0.013} & 1 &0.103 &0.213 &0.068 &0.565 & 79
& 66.6  \\
V & 528 & 10 &0.610 &0.491 & 2 & {\bf0.387} &0.480 &0.409 &0.508 & 1 &
19.5  \\
\hline
\end{tabular*}
\tabnotetext[*]{TT1}{\%Imp: Percent relative improvement of customized training to standard training.}
\end{table}

Random forests achieve the lowest error on 8 of the 16
data sets, the most of any method. But the method that achieves the
lowest error
secondmost often is customized training, on 7 of the 16 data sets, and
customized training beats standard training on 11 data sets, with standard
training coming out on top for only 2 data sets. We do not expect customized
training to provide value on
all data sets, but through cross-validation, we can often identify data
sets for
which standard training is better, meaning that $G = 1$ is chosen through
cross-validation. The point of this exercise is not to show that customized
training is superior to the other methods but rather to show that,
despite its
simplicity, it is at least competitive with the other methods.

% latex table generated in R 3.0.2 by xtable 1.7-3 package
% Tue Apr 29 14:20:28 2014

Table~\ref{tab-rejections} shows all of the rejections that customized training
makes on
the 16 data sets, for any value of $G$ (not just the values of $G$
chosen by
cross-validation). For two of the data sets ({\tt LSVT Voice Rehabilitation}
and {\tt Parkinsons}), it is clear that the rejections are just
artifacts of
using a $G$ that is too large relative to the training sample size $n$.
Such a
$G$ is not chosen by cross-validation. However, in the other data sets,
{\tt Steel Plates Faults} and {\tt First-order theorem proving}, rejections
occur for moderate values of $G$. It
seems that this rejection is appropriate because the standard training method
leads to an error for each test point which is rejected. Overall, we observe
that rejections are rare.

%\begin{table}[h]
%\caption[Dataset used from UCI Machine Learning Repository]{Datasets
%from UCI
%Machine Learning Repository \citet{UCI} used in Section
%\ref{sec-battery}.}
%\label{tab-abbreviations}
%\begin{center}
%\begin{tabular}{ll}
%Abbrv. & Data set name\\
%\hline
%BS & Balance scale\\
%BCW & Breast cancer Wisconsin (diagnostic)\\
%C & Chess (king-rook vs king-pawn)\\
%CMC & Contraceptive method choice\\
%F & Fertility\\
%FOTP & First-order theorem proving\\
%LSVT & LSVT voice rehabilitation\\
%M & Mushroom\\
%ORHD & Optical recognition of handwritten digits\\
%P & Parkinsons\\
%QSAR & QSAR biodegration\\
%S & Seeds\\
%SPF & Steel plates faults\\
%TAE & Teaching assistant evaluation\\
%UKM & User knowledge modeling\\
%V & Vowel
%\end{tabular}
%\end{center}
%\end{table}

%s6 #&#
\section{Discussion}
\label{sec-discussion}

The idea behind customized training is simple: for each subset of the test
data, identify a customized subset of the training data that is close
to this
subset and use this data to train a customized model. We proposed two
different clustering methods for finding the customized training sets
and used
$\ell_1$-regularized methods for training the models. Local learning
has been
used in the transductive setting but not in such a parsimonious, interpretable
way. Customized training has the potential to uncover hidden regimes in the
data and leverage this discovery to make better predictions. It may be that
some classes are over-represented in a cluster, and fitting a model in this
cluster effectively customizes the prior to reflect this over-representation.
Our results demonstrate superior performance of customized training over
standard training on the mass-spectrometric imaging data set of gastric cancer
surgical resection margins, in terms of discrimination between cancer and
normal cells. Our approach also suggests the possibility of two
subclasses of
cancer, consistent with a speculation raised by our medical collaborators.

%
%t5 #&#
\begin{table}[t]
\tabcolsep=0pt
\caption{A listing of all data sets from Section~\protect\ref{sec-battery} for which $K$-$\mathrm{CT}_J$ makes a rejection for some $K$. The
error rates in the last two columns refer to the error rate of standard training}\label{tab-rejections}
\begin{tabular*}{\tablewidth}{@{\extracolsep{\fill}}@{}lccd{1.2}c@{}}
\hline
& & & \multicolumn{1}{c}{\textbf{Error rate on}} & \multicolumn{1}{c}{\textbf{Error rate}}\\
\textbf{Data set} & \textbf{Method} & \textbf{Rejections} & \multicolumn{1}{c}{\textbf{rejections}} & \textbf{overall}\\
\hline
First-order theorem proving & 3-$\mathrm{CT}_J$ & 3 & 1 &0.518\\
& 5-$\mathrm{CT}_J$ & 3 & 1\\
& 10-$\mathrm{CT}_J$ & 3 & 1
\\[3pt]
LSVT Voice Rehabilitation & 10-$\mathrm{CT}_J$ & 2 & 0.5 &0.142
\\[3pt]
Parkinsons & 10-$\mathrm{CT}_J$ & 4 & 0.25 &0.154
\\[3pt]
Steel Plates Faults & 3-$\mathrm{CT}_J$ & 1 & 1 &0.294\\
& 5-$\mathrm{CT}_J$ & 1 & 1\\
& 10-$\mathrm{CT}_J$ & 1 & 1\\
\hline
\end{tabular*}
\end{table}

In this paper we focused on customized training with $\ell_1$-regularized
methods for the sake of interpretability, but, in principle, any supervised
learning method may be used, which is an area for future work. Another area
of future work is the use of different clustering techniques. We use
hierarchical clustering, but there may be value in other methods, such as
prototype clustering [\citet{BienTibshirani11}]. Simulations in
Section~\ref{sec-simulation} show that the method can struggle in the
high-dimensional
setting, so it may be worthwhile to consider sparse clustering
[\citet{WittenTibshirani10}].

\section*{Acknowledgments}
The authors would like to thank Dr. Livia Eberlin and Professor Richard Zare
for allowing them to use the gastric cancer data and are grateful to an
Editor and an Associate Editor for helpful comments that led to improvements
to this work.

%\begin{appendix}
%\section{}
%\end{appendix}

% zodis "Acknowledgments" paliekamas pagal autoriu
%\section*{Acknowledgments}

%\begin{supplement}[id=suppA]
%\sname{Supplement A}
%\stitle{}
%\slink[doi]{10.1214/00-AOASXXXXSUPP} %[doi,text={...}] - jei reikia
%suskaldyti doi
%\sdatatype{.pdf}
%\sfilename{aoasXXXX\_supp.pdf}
%\sdescription{}
%\end{supplement}

% imsref loaded by linak, 2015-10-07 14:55:46
%

\printaddresses

\begin{thebibliography}{27}
% pybtex-1.44. Style name=ims, version=2.92, label_style=nameyear,
%sorting_style=complex, cfg=None, language=None.
%b1 ###
%b1 #&#
\bibitem[\protect\citeauthoryear{Bache and Lichman}{2013}]{UCI}
%
\begin{bmisc}[author]
\bauthor{\bsnm{Bache},~\bfnm{K.}\binits{K.}} \AND
\bauthor{\bsnm{Lichman},~\bfnm{M.}\binits{M.}}
(\byear{2013}).
\bhowpublished{UCI machine learning repository.
Univ. California  Irvine School of Information and Computer Science,  Irvine, CA}.
\end{bmisc}
%

\bptok{imsref}%
% NOT OUTPUTTED:
% sortkey = Bache(2013
\endbibitem

%b2 ###
%b2 #&#
\bibitem[\protect\citeauthoryear{Bien and Tibshirani}{2011}]{BienTibshirani11}
%
\begin{barticle}[mr]
\bauthor{\bsnm{Bien},~\bfnm{Jacob}\binits{J.}} \AND
\bauthor{\bsnm{Tibshirani},~\bfnm{Robert}\binits{R.}}
(\byear{2011}).
\btitle{Hierarchical clustering with prototypes via minimax linkage}.
\bjournal{J. Amer. Statist. Assoc.}
\bvolume{106}
\bpages{1075--1084}.
\bid{doi={10.1198/jasa.2011.tm10183}, issn={0162-1459}, mr={2894765}}
\end{barticle}
%

\bptok{imsref}%
% NOT OUTPUTTED:
% number = 495
% doi = http://dx.doi.org/10.1198/jasa.2011.tm10183
% coden = JSTNAL
% fjournal = Journal of the American Statistical Association
\endbibitem

%b3 ###
%b3 #&#
\bibitem[\protect\citeauthoryear{Bottou and Vapnik}{1992}]{BottouVapnik92}
%
\begin{barticle}[author]
\bauthor{\bsnm{Bottou},~\bfnm{L{\'e}on}\binits{L.}} \AND
\bauthor{\bsnm{Vapnik},~\bfnm{Vladimir}\binits{V.}}
(\byear{1992}).
\btitle{Local learning algorithms}.
\bjournal{Neural Comput.}
\bvolume{4}
\bpages{888--900}.
\end{barticle}
%

\bptok{imsref}%
\endbibitem

%b4 ###
%b4 #&#
\bibitem[\protect\citeauthoryear{Cortes and Mohri}{2007}]{CortesMohri07}
%
\begin{binproceedings}[author]
\bauthor{\bsnm{Cortes},~\bfnm{Corinna}\binits{C.}} \AND
\bauthor{\bsnm{Mohri},~\bfnm{Mehryar}\binits{M.}}
(\byear{2007}).
\btitle{On transductive regression}.
In \bbooktitle{Advances in Neural Information Processing Systems 19}.
\blocation{Vancouver, BC, Canada}.
\end{binproceedings}
%

\bptok{imsref}%
\endbibitem

%b5 ###
%b5 #&#
\bibitem[\protect\citeauthoryear{Eberlin et~al.}{2014}]{Eberlin.etal14}
%
\begin{barticle}[author]
\bauthor{\bsnm{Eberlin},~\bfnm{L~S.}\binits{L.~S.}},
\bauthor{\bsnm{Tibshirani},~\bfnm{Robert~J.}\binits{R.~J.}},
\bauthor{\bsnm{Zhang},~\bfnm{J.}\binits{J.}},
\bauthor{\bsnm{Longacre},~\bfnm{T~A.}\binits{T.~A.}},
\bauthor{\bsnm{Berry},~\bfnm{G~J.}\binits{G.~J.}},
\bauthor{\bsnm{Bingham},~\bfnm{D~B.}\binits{D.~B.}},
\bauthor{\bsnm{Norton},~\bfnm{J~A.}\binits{J.~A.}},
\bauthor{\bsnm{Zare},~\bfnm{R~N.}\binits{R.~N.}} \AND
\bauthor{\bsnm{Poultsides},~\bfnm{G~A.}\binits{G.~A.}}
(\byear{2014}).
\btitle{{Molecular assessment of surgical-resection margins of gastric
cancer by mass-spectrometric imaging}}.
\bjournal{Proc. Natl. Acad. Sci. USA}
\bvolume{111}
\bpages{2436--2441}.
\end{barticle}
%

\bptok{imsref}%
\endbibitem

%b6 ###
%b6 #&#
\bibitem[\protect\citeauthoryear{Fu, Robles-Kelly and Zhou}{2010}]{Fu.etal10}
%
\begin{barticle}[author]
\bauthor{\bsnm{Fu},~\bfnm{Zhouyu}\binits{Z.}},
\bauthor{\bsnm{Robles-Kelly},~\bfnm{Antonio}\binits{A.}} \AND
\bauthor{\bsnm{Zhou},~\bfnm{Jun}\binits{J.}}
(\byear{2010}).
\btitle{{Mixing linear SVMs for nonlinear classification}}.
\bjournal{IEEE Trans. Neural Netw.}
\bvolume{21}
\bpages{1963--1975}.
\end{barticle}
%

\bptok{imsref}%
\endbibitem

%b7 ###
%b7 #&#
\bibitem[\protect\citeauthoryear{Gil et~al.}{2012}]{Gil.etal12}
%
\begin{barticle}[author]
\bauthor{\bsnm{Gil},~\bfnm{David}\binits{D.}},
\bauthor{\bsnm{Girela},~\bfnm{Jose~Luis}\binits{J.~L.}},
\bauthor{\bsnm{De~Juan},~\bfnm{Joaquin}\binits{J.}},
\bauthor{\bsnm{Gomez-Torres},~\bfnm{M~Jose}\binits{M.~J.}} \AND
\bauthor{\bsnm{Johnsson},~\bfnm{Magnus}\binits{M.}}
(\byear{2012}).
\btitle{{Predicting seminal quality with artificial intelligence methods}}.
\bjournal{Expert Syst. Appl.}
\bvolume{39}
\bpages{12564--12573}.
\end{barticle}
%

\bptok{imsref}%
\endbibitem

%b8 ###
%b8 #&#
\bibitem[\protect\citeauthoryear{Gu and Han}{2013}]{GuHan13}
%
\begin{binproceedings}[author]
\bauthor{\bsnm{Gu},~\bfnm{Quanquan}\binits{Q.}} \AND
\bauthor{\bsnm{Han},~\bfnm{Jiawei}\binits{J.}}
(\byear{2013}).
\btitle{Clustered support vector machines}.
In \bbooktitle{Proceedings of the 16th International Conference on
Artificial Intelligence and Statistics}
\bpages{307--315}.
\blocation{Scottsdale, AZ}.
\bnote{Available at DOI:\doiurl{10.1186/1475-925X-6-23}.}
\end{binproceedings}
%

\bptok{imsref}%
\endbibitem

%b9 ###
%b9 #&#
\bibitem[\protect\citeauthoryear{Hamburg and Collins}{2010}]{HamburgCollins10}
%
\begin{barticle}[pbm]
\bauthor{\bsnm{Hamburg},~\bfnm{Margaret~A.}\binits{M.~A.}} \AND
\bauthor{\bsnm{Collins},~\bfnm{Francis~S.}\binits{F.~S.}}
(\byear{2010}).
\btitle{The path to personalized medicine}.
\bjournal{N. Engl. J. Med.}
\bvolume{363}
\bpages{301--304}.
\bid{doi={10.1056/NEJMp1006304}, issn={1533-4406}, pii={NEJMp1006304},
pmid={20551152}}
\end{barticle}
%

\bptok{imsref}%
% NOT OUTPUTTED:
% number = 4
% fjournal = The New England journal of medicine
\endbibitem

%b10 ###
%b10 #&#
\bibitem[\protect\citeauthoryear{Hastie, Tibshirani and
Friedman}{2009}]{Hastie.etal09}
%
\begin{bbook}[mr]
\bauthor{\bsnm{Hastie},~\bfnm{Trevor}\binits{T.}},
\bauthor{\bsnm{Tibshirani},~\bfnm{Robert}\binits{R.}} \AND
\bauthor{\bsnm{Friedman},~\bfnm{Jerome}\binits{J.}}
(\byear{2009}).
\btitle{The Elements of Statistical Learning: Data Mining, Inference,
and Prediction},
\bedition{2nd} ed.
\bpublisher{Springer},
\blocation{New York}.
\bid{doi={10.1007/978-0-387-84858-7}, mr={2722294}}
\end{bbook}
%

\bptok{imsref}%
% NOT OUTPUTTED:
% doi = http://dx.doi.org/10.1007/978-0-387-84858-7
% isbn = 978-0-387-84857-0
% fpage = xxii+745
\endbibitem

%b11 ###
%b11 #&#
\bibitem[\protect\citeauthoryear{Jordan and Jacobs}{1994}]{JordanJacobs94}
%
\begin{barticle}[author]
\bauthor{\bsnm{Jordan},~\bfnm{Michael~I.}\binits{M.~I.}} \AND
\bauthor{\bsnm{Jacobs},~\bfnm{Robert~A.}\binits{R.~A.}}
(\byear{1994}).
\btitle{Hierarchical mixtures of experts and the EM algorithm}.
\bjournal{Neural Comput.}
\bvolume{6}
\bpages{181--214}.
\end{barticle}
%

\bptok{imsref}%
\endbibitem

%b12 ###
%b12 #&#
\bibitem[\protect\citeauthoryear{Kahraman, Sagiroglu and
Colak}{2013}]{Kahraman.etal13}
%
\begin{barticle}[author]
\bauthor{\bsnm{Kahraman},~\bfnm{H~Tolga}\binits{H.~T.}},
\bauthor{\bsnm{Sagiroglu},~\bfnm{Seref}\binits{S.}} \AND
\bauthor{\bsnm{Colak},~\bfnm{Ilhami}\binits{I.}}
(\byear{2013}).
\btitle{The development of intuitive knowledge classifier and the
modeling of domain dependent data}.
\bjournal{Knowledge-Based Systems}
\bvolume{37}
\bpages{283--295}.
\end{barticle}
%

\bptok{imsref}%
\endbibitem

%b13 ###
%b13 #&#
\bibitem[\protect\citeauthoryear{Ladicky and Torr}{2011}]{LadickyTorr11}
%
\begin{binproceedings}[author]
\bauthor{\bsnm{Ladicky},~\bfnm{Lubor}\binits{L.}} \AND
\bauthor{\bsnm{Torr},~\bfnm{Phillip}\binits{P.}}
(\byear{2011}).
\btitle{{Locally linear support vector machines}}.
In \bbooktitle{Proceedings of the 28th International Conference on
Machine Learning}
\bpages{985--992}.
\blocation{Bellevue, WA}.
\end{binproceedings}
%

\bptok{imsref}%
\endbibitem

%b14 ###
%b14 #&#
\bibitem[\protect\citeauthoryear{Little et~al.}{2007}]{Little.etal07}
%
\begin{barticle}[author]
\bauthor{\bsnm{Little},~\bfnm{Max~A.}\binits{M.~A.}},
\bauthor{\bsnm{McSharry},~\bfnm{Patrick~E.}\binits{P.~E.}},
\bauthor{\bsnm{Roberts},~\bfnm{Stephen~J.}\binits{S.~J.}},
\bauthor{\bsnm{Costello},~\bfnm{Declan~AE}\binits{D.~A.}} \AND
\bauthor{\bsnm{Moroz},~\bfnm{Irene~M.}\binits{I.~M.}}
(\byear{2007}).
\btitle{Exploiting nonlinear recurrence and fractal scaling properties
for voice disorder detection}.
\bjournal{Biomed. Eng. Online}
\bvolume{6}
\bpages{23}.
\end{barticle}
%

\bptok{imsref}%
\endbibitem

%b15 ###
%b15 #&#
\bibitem[\protect\citeauthoryear{Ma}{2012}]{Ma12}
%
\begin{bmisc}[author]
\bauthor{\bsnm{Ma},~\bfnm{Tian~Min}\binits{T.~M.}}
(\byear{2012}).
\bhowpublished{Local and personalised modelling for renal medical
decision support system.
Ph.D. thesis, Auckland Univ.  Technology}.
\end{bmisc}
%

\bptok{imsref}%
% NOT OUTPUTTED:
% publisher = Auckland Univ. Technology
\endbibitem

%b16 ###
%b16 #&#
\bibitem[\protect\citeauthoryear{Mansouri et~al.}{2013}]{Mansouri.etal13}
%
\begin{barticle}[pbm]
\bauthor{\bsnm{Mansouri},~\bfnm{Kamel}\binits{K.}},
\bauthor{\bsnm{Ringsted},~\bfnm{Tine}\binits{T.}},
\bauthor{\bsnm{Ballabio},~\bfnm{Davide}\binits{D.}},
\bauthor{\bsnm{Todeschini},~\bfnm{Roberto}\binits{R.}} \AND
\bauthor{\bsnm{Consonni},~\bfnm{Viviana}\binits{V.}}
(\byear{2013}).
\btitle{Quantitative structure-activity relationship models for ready
biodegradability of chemicals}.
\bjournal{J.~Chem. Inf. Model.}
\bvolume{53}
\bpages{867--878}.
\bid{doi={10.1021/ci4000213}, issn={1549-960X}, pmid={23469921}}
\end{barticle}
%

\bptok{imsref}%
% NOT OUTPUTTED:
% number = 4
% fjournal = Journal of chemical information and modeling
\endbibitem

%b17 ###
%b17 #&#
\bibitem[\protect\citeauthoryear{Shahbaba and Neal}{2009}]{ShahbabaNeal09}
%
\begin{barticle}[mr]
\bauthor{\bsnm{Shahbaba},~\bfnm{Babak}\binits{B.}} \AND
\bauthor{\bsnm{Neal},~\bfnm{Radford}\binits{R.}}
(\byear{2009}).
\btitle{Nonlinear models using {D}irichlet process mixtures}.
\bjournal{J. Mach. Learn. Res.}
\bvolume{10}
\bpages{1829--1850}.
\bid{issn={1532-4435}, mr={2540778}}
\end{barticle}
%

\bptok{imsref}%
% NOT OUTPUTTED:
% fjournal = Journal of Machine Learning Research (JMLR)
\endbibitem

%b18 ###
%b18 #&#
\bibitem[\protect\citeauthoryear{Tibshirani}{1996}]{Tibshirani96}
%
\begin{barticle}[mr]
\bauthor{\bsnm{Tibshirani},~\bfnm{Robert}\binits{R.}}
(\byear{1996}).
\btitle{Regression shrinkage and selection via the lasso}.
\bjournal{J. Roy. Statist. Soc. Ser. B}
\bvolume{58}
\bpages{267--288}.
\bid{issn={0035-9246}, mr={1379242}}
\end{barticle}
%

\bptok{imsref}%
% NOT OUTPUTTED:
% url =
%http://links.jstor.org/sici?sici=0035-9246(1996)58:1<267:RSASVT>2.0.CO;2-G&origin=MSN
% number = 1
% coden = JSTBAJ
% fjournal = Journal of the Royal Statistical Society. Series B.
%Methodological
\endbibitem

%b19 ###
%b19 #&#
\bibitem[\protect\citeauthoryear{Torgo and DaCosta}{2003}]{TorgoDaCosta03}
%
\begin{barticle}[author]
\bauthor{\bsnm{Torgo},~\bfnm{Luis}\binits{L.}} \AND
\bauthor{\bsnm{DaCosta},~\bfnm{Joaquim~Pinto}\binits{J.~P.}}
(\byear{2003}).
\btitle{Clustered partial linear regression}.
\bjournal{Mach. Learn.}
\bvolume{50}
\bpages{303--319}.
\end{barticle}
%

\bptok{imsref}%
\endbibitem

%b20 ###
%b20 #&#
\bibitem[\protect\citeauthoryear{Tsanas et~al.}{2014}]{Tsanas.etal13}
%
\begin{barticle}[pbm]
\bauthor{\bsnm{Tsanas},~\bfnm{Athanasios}\binits{A.}},
\bauthor{\bsnm{Little},~\bfnm{Max~A.}\binits{M.~A.}},
\bauthor{\bsnm{Fox},~\bfnm{Cynthia}\binits{C.}} \AND
\bauthor{\bsnm{Ramig},~\bfnm{Lorraine~O.}\binits{L.~O.}}
(\byear{2014}).
\btitle{Objective automatic assessment of rehabilitative speech
treatment in Parkinson's disease}.
\bjournal{IEEE Trans. Neural Syst. Rehabil. Eng.}
\bvolume{22}
\bpages{181--190}.
\bid{doi={10.1109/TNSRE.2013.2293575}, issn={1558-0210}, pmid={26271131}}
\end{barticle}
%

\bptok{imsref}%
% NOT OUTPUTTED:
% number = 1
% fjournal = IEEE transactions on neural systems and rehabilitation
%engineering : a publication of the IEEE Engineering in Medicine and
%Biology Society
\endbibitem

%b21 ###
%b21 #&#
\bibitem[\protect\citeauthoryear{Witten and
Tibshirani}{2010}]{WittenTibshirani10}
%
\begin{barticle}[mr]
\bauthor{\bsnm{Witten},~\bfnm{Daniela~M.}\binits{D.~M.}} \AND
\bauthor{\bsnm{Tibshirani},~\bfnm{Robert}\binits{R.}}
(\byear{2010}).
\btitle{A framework for feature selection in clustering}.
\bjournal{J.~Amer. Statist. Assoc.}
\bvolume{105}
\bpages{713--726}.
\bid{doi={10.1198/jasa.2010.tm09415}, issn={0162-1459}, mr={2724855}}
\end{barticle}
%

\bptok{imsref}%
% NOT OUTPUTTED:
% number = 490
% doi = http://dx.doi.org/10.1198/jasa.2010.tm09415
% coden = JSTNAL
% fjournal = Journal of the American Statistical Association
\endbibitem

%b22 ###
%b22 #&#
\bibitem[\protect\citeauthoryear{World Health Organization}{2013}]{WHO13}
%
\begin{bmisc}[author]
\bauthor{\bsnm{World Health Organization}}
(\byear{2013}).
\bhowpublished{Cancer. WHO Fact Sheet No. 297.
Available at \surl{http://www.who.int/mediacentre/factsheets/fs297/en/index.html}}.
\end{bmisc}
%

\bptok{imsref}%
% NOT OUTPUTTED:
% url = Available at
%http://www.who.int/mediacentre/factsheets/fs297/en/index.html
% sortkey = World(2013
\endbibitem

%b23 ###
%b23 #&#
\bibitem[\protect\citeauthoryear{Wu and Sch{\"o}lkopf}{2007}]{WuSchoelkopf07}
%
\begin{binproceedings}[author]
\bauthor{\bsnm{Wu},~\bfnm{Mingrui}\binits{M.}} \AND
\bauthor{\bsnm{Sch{\"o}lkopf},~\bfnm{Bernhard}\binits{B.}}
(\byear{2007}).
\btitle{{Transductive classification via local learning regularization}}.
In \bbooktitle{Proceedings of the 11th International Conference on
Artificial Intelligence and Statistics}
\bpages{628--635}.
\blocation{San Juan, Puerto Rico}.
\end{binproceedings}
%

\bptok{imsref}%
\endbibitem

%b24 ###
%b24 #&#
\bibitem[\protect\citeauthoryear{Yu, Zhang and Gong}{2009}]{Yu.etal09}
%
\begin{barticle}[author]
\bauthor{\bsnm{Yu},~\bfnm{Kai}\binits{K.}},
\bauthor{\bsnm{Zhang},~\bfnm{Tong}\binits{T.}} \AND
\bauthor{\bsnm{Gong},~\bfnm{Yihong}\binits{Y.}}
(\byear{2009}).
\btitle{Nonlinear learning using local coordinate coding}.
\bjournal{Adv. Neural Inf. Process. Syst.}
\bvolume{21}
\bpages{2223--2231}.
\end{barticle}
%

\bptok{imsref}%
\endbibitem

%b25 ###
%b25 #&#
\bibitem[\protect\citeauthoryear{Zhou et~al.}{2004}]{Zhou.etal04}
%
\begin{barticle}[author]
\bauthor{\bsnm{Zhou},~\bfnm{Dengyong}\binits{D.}},
\bauthor{\bsnm{Bousquet},~\bfnm{Olivier}\binits{O.}},
\bauthor{\bsnm{Lal},~\bfnm{Thomas~Navin}\binits{T.~N.}},
\bauthor{\bsnm{Weston},~\bfnm{Jason}\binits{J.}} \AND
\bauthor{\bsnm{Sch{\"o}lkopf},~\bfnm{Bernhard}\binits{B.}}
(\byear{2004}).
\btitle{Learning with local and global consistency}.
\bjournal{Adv. Neural Inf. Process. Syst.}
\bvolume{16}
\bpages{321--328}.
\end{barticle}
%

\bptok{imsref}%
\endbibitem

%b26 ###
%b26 #&#
\bibitem[\protect\citeauthoryear{Zhu}{2007}]{Zhu07}
%
\begin{bmisc}[author]
\bauthor{\bsnm{Zhu},~\bfnm{Xiaojin}\binits{X.}}
(\byear{2007}).
\btitle{Semi-supervised learning literature survey.
Technical Report No. 1530, Dept. Computer Science,
Univ. Wisconsin-Madison, Madison, WI}.
\end{bmisc}
%

\bptok{imsref}%
% NOT OUTPUTTED:
% publisher = Univ. Wisconsin-Madison Department of Computer Science
\endbibitem

%b27 ###
%b27 #&#
\bibitem[\protect\citeauthoryear{Zhu, Chen and Xing}{2011}]{Zhu.etal11}
%
\begin{binproceedings}[author]
\bauthor{\bsnm{Zhu},~\bfnm{Jun}\binits{J.}},
\bauthor{\bsnm{Chen},~\bfnm{Ning}\binits{N.}} \AND
\bauthor{\bsnm{Xing},~\bfnm{Eric~P.}\binits{E.~P.}}
(\byear{2011}).
\btitle{Infinite SVM: A Dirichlet process mixture of large-margin
kernel machines}.
In \bbooktitle{Proceedings of the 28th International Conference on Machine Learning}
\bpages{617--624}.
\blocation{Bellevue, WA}.
\end{binproceedings}
%

\bptok{imsref}%
\endbibitem
\end{thebibliography}
\end{document}